# Enhancement of dielectric properties of lead-free BCZT ferroelectric ceramics by grain size engineering

Z. Hanani [a,b], D. Mezzane [a], M. Amjoud [a], S. Fourcade [b], A.G Razumnaya [c,d], I.A. Luk'yanchuk [d,e] and M. Gouné [b,*]

[a] LMCN, Cadi Ayyad University, Marrakesh, 40000, Morocco

[b] ICMCB, University of Bordeaux, Pessac, 33600, France

[c] Faculty of Physics, Southern Federal University, Rostov-on-Don, 344090, Russia

[d] LPMC, University of Picardy Jules Verne, Amiens, 80039, France

[e] Landau Institute for Theoretical Physics, Moscow, 119334, Russia

[*] Corresponding author: Mohamed.Goune@icmcb.cnrs.fr

## Abstract

Lead-free $Ba_{0.85}Ca_{0.15}Ti_{0.9}Zr_{0.1}O_3$ (BCZT) ceramics had attracted much attention for the fabrication of microelectronic devices by virtue of their excellent dielectric, ferroelectric and piezoelectric properties. To understand the effects of both mean grain size and grain size distribution on the dielectric properties of lead-free $Ba_{0.85}Ca_{0.15}Ti_{0.9}Zr_{0.1}O_3$ (BCZT) ferroelectric relaxors, an original method was proposed. It is based on the surfactant-assisted solvothermal processing coupled with low-temperature conventional sintering at 1250 °C. In this way, three highly dense BCZT with different mean grain size and dissimilar grain size distribution were designed. A significant increase of dielectric properties was obtained by a control of grain size and densification process. The dielectric constants measured were ranged from 5370 to 9646 and the dielectric loss was enhanced by 70%. Surprisingly, it was evidenced that there is unequivocal link between mean grain size with dielectric properties. Indeed, it was found that the presence of high density of refined grains leads to an improvement of dielectric properties due to an enhancement of densification. This work may provide a new strategy to design ferroelectric materials with enhanced properties.

# 1. Introduction

Quite recently, considerable attention has been paid to develop new lead-free ferroelectric ceramics due to their non-toxic behavior [1,2]. Among this class of eco-friendly ceramics, barium titanate ($BaTiO_3$) with perovskite structure ($ABO_3$) was extensively investigated by dint of its excellent dielectric and piezoelectric properties [3,4]. However, as compared with lead-based perovskite ceramics such as (PZT), pure $BaTiO_3$ exhibits relatively low dielectric constant ($\varepsilon_r$) piezoelectric coefficient ($d_{33}$) [5]. To overcome this drawback, many efforts were concentrating on doping strategy to modify barium titanate [6–8]. For instance, $Ca^{2+}$ and $Zr^{4+}$ can be introduced into the crystal lattice of $BaTiO_3$, in order to replace $Ba^{2+}$ and $Ti^{4+}$, respectively [9–11]. W. Liu and X. Ren [12] reported non-Pb ceramics of $0.5Ba(Zr_{0.2}Ti_{0.8})O_3–0.5(Ba_{0.7}Ca_{0.3})TiO_3$ with extremely high $\varepsilon_r$ of ≈ 18 000 and $d_{33}$ of 620 pC/N at the Morphotropic Phase Boundary (MPB). Inevitably, the basic approach to achieve high dielectric and ferroelectric properties relay on placing the composition of the material to the proximity of the MPB [13,14].

There is an unambiguous relationship between microstructure and properties of any material directly influence its properties. Thus, through a wide understanding of the vital role which plays grain size, it can be easily to enhance the ferroelectric and dielectric properties of lead-free ceramics at the Morphotropic Phase Boundary (MPB) [15–17]. H. L. Sun et al. [5] reported a correlation of grain size, phase transition and piezoelectric properties in $Ba_{0.85}Ca_{0.15}Ti_{0.90}Zr_{0.10}O_3$ ceramics prepared via conventional solid-state reaction and at sintering temperature rising from 1320 to 1560 °C and a dwell time of 2 h. They suggested that the ceramic sintered at 1480 °C/2h possesses the optimum piezoelectricity, which was attributed to the largest grain size together with more noticeable rhombohedral–tetragonal phase transition near room temperature [5]. A detailed investigation of a wide range of processing factors, including $Ba(Zr,Ti)O_3/(Ba,Ca)TiO_3$ ratio, sintering conditions, particle size of the calcined ceramic powder, structure and microstructure (e.g. phase, lattice parameters, density and grain size), and their effect on the piezoelectric properties has been provided by Y. Bai et al. [18]. This research group revealed the importance of the degree of tetragonality (c/a) of calcined ceramic powder, and grain size growth and densification process in sintered ceramics to enhance the dielectric properties of BCZT systems prepared by conventional solid-state reaction [18]. Y. Tan et al. [19] unfolded clearly the effects of grains size in $BaTiO_3$ ferroelectric ceramics prepared by conventional sintering (CS) and spark plasma sintering (SPS) using micro- and nano-sized powders. They demonstrated that the grain size effect on the dielectric permittivity is nearly independent of the starting powder used and sintering method. The maximum dielectric constant was achieved near 1 μm grain size by optimum density and mobility of 90° domain walls in all the $BaTiO_3$ ceramics studied [19].

To date, to correlate the grain size and dielectric properties of $BaTiO_3$-based materials a large and growing body of literature are focusing on sintering process to elucidate the effect of grain size on dielectric properties [20]. Consequently, this can be reached by varying sintering temperature [21,22] and dwell time [5,23]. Although, a key limitation of this approach is that a higher sintering temperature is required for better comparison [18,24,25]. The aim of this study is to shine new light for an effective comparison of the effects of grain size distribution and dielectric properties of $Ba_{0.85}Ca_{0.15}Ti_{0.9}Zr_{0.1}O_3$ (BCZT) ceramics with different grain sizes at fixed sintering temperature and dwell time. This can be achieved using solvothermal processing via the addition of two types of surfactants to easily control grain size distribution of the final product [26,27]. Consequently, a logical correlation between grain size and dielectric properties can be unfolded at low and fixed sintering temperature and time.

## 2. Experimental

*2.1. BCZT Powders synthesis*

B-0, B-CTAB and BCZT-SDS pure crystalline powders were obtained through single-step solvothermal synthesis, by controlled reaction at room temperature between calcium nitrate tetrahydrate, zirconium n-propoxide, barium acetate and titanium $^{(IV)}$ isopropoxide. All the reagents have analytical grade and were purchased from Sigma-Aldrich. First, an appropriate amount of barium acetate was dissolved in glacial acetic acid. Calcium nitrate tetrahydrate was dissolved in 2-ethoxyethanol. Second, the two solutions were mixed in a 50-ml round-bottom flask equipped with a magnetic stirrer. Third, titanium [IV] isopropoxide and zirconium n-propoxide were added quickly to the reaction medium according to the stoichiometric formula $Ba_{0.85}Ca_{0.15}Ti_{0.9}Zr_{0.1}O_3$.

A fixed concentration of CTAB (Cetyltrimethylammonium bromide, $C_{19}H_{42}BrN$) or SDS (sodium dodecyl sulfate, $NaC_{12}H_{25}SO_4$) was added to each solution with an additional 1 hour of continuous stirring at room temperature. The obtained mixtures were transferred separately to 30-ml Teflon-lined stainless-steel autoclaves at 180 °C in an oven for 12 h. After the reaction was completed, the sealed autoclaves were cooled naturally to room temperature. The resulting white precipitates were collected by centrifugation at 12 000 rpm for 10 min, and washed several times with deionized water and ethanol. Then, the final products were dried at 100 °C for 12 hours and calcined at 1000 °C for 4 hours to get fine powders. For electrical measurements, B-0, B-CTAB and B-SDS calcined powders were uni-axially pressed into pellets of diameter about 6 mm and thickness about 1 mm, and then sintered at 1250 °C/10h. For comparison, B-0 sample was synthesized under the same conditions but without the addition of any surfactant.

*2.2. Characterization*

Crystalline structure of B-0, B-CTAB and B-SDS was performed by X-ray diffraction (XRD, Panalytical X-Pert Pro). The measurement has been done at room temperature employing a step angle of 0.017° in the 2θ range from 10 to 80° using a Cu-$K_\alpha$ radiation (λ ~ 1.540598 Å). The resulting microstructures were analyzed using a Scanning Electron Microscope (SEM, Tescan VEGA-3). The density of the sintered ceramics was measured by Archimedes method using deionized water as a medium. A precision LCR Meter (HP 4284A, 20 Hz to 1 MHz) was used to measure the dielectric properties of BCZT sintered pellets electroded with silver paste in the frequency range 20 Hz to 1 MHz.

# 3. Results and discussion

*3.1. Phase characterization of nanopowders*

The XRD patterns obtained at room temperature for B-0, B-CTAB and B-SDS powders calcined at 1000 °C are illustrated in Fig. 1. All BCZT powders were formed with a pure perovskite phase, without any secondary phase peaks. The peak splitting between 2θ ≈ 44 - 46° indicates the presence of the tetragonal phase (Fig. 1(b)). However, the formation of the triplet around 2θ ≈ 66° suggests the presence of the orthorhombic phase (Fig. 1(c)). The co-existence of tetragonal and orthorhombic phases at room temperature was also reported by others researchers [12,28–30].

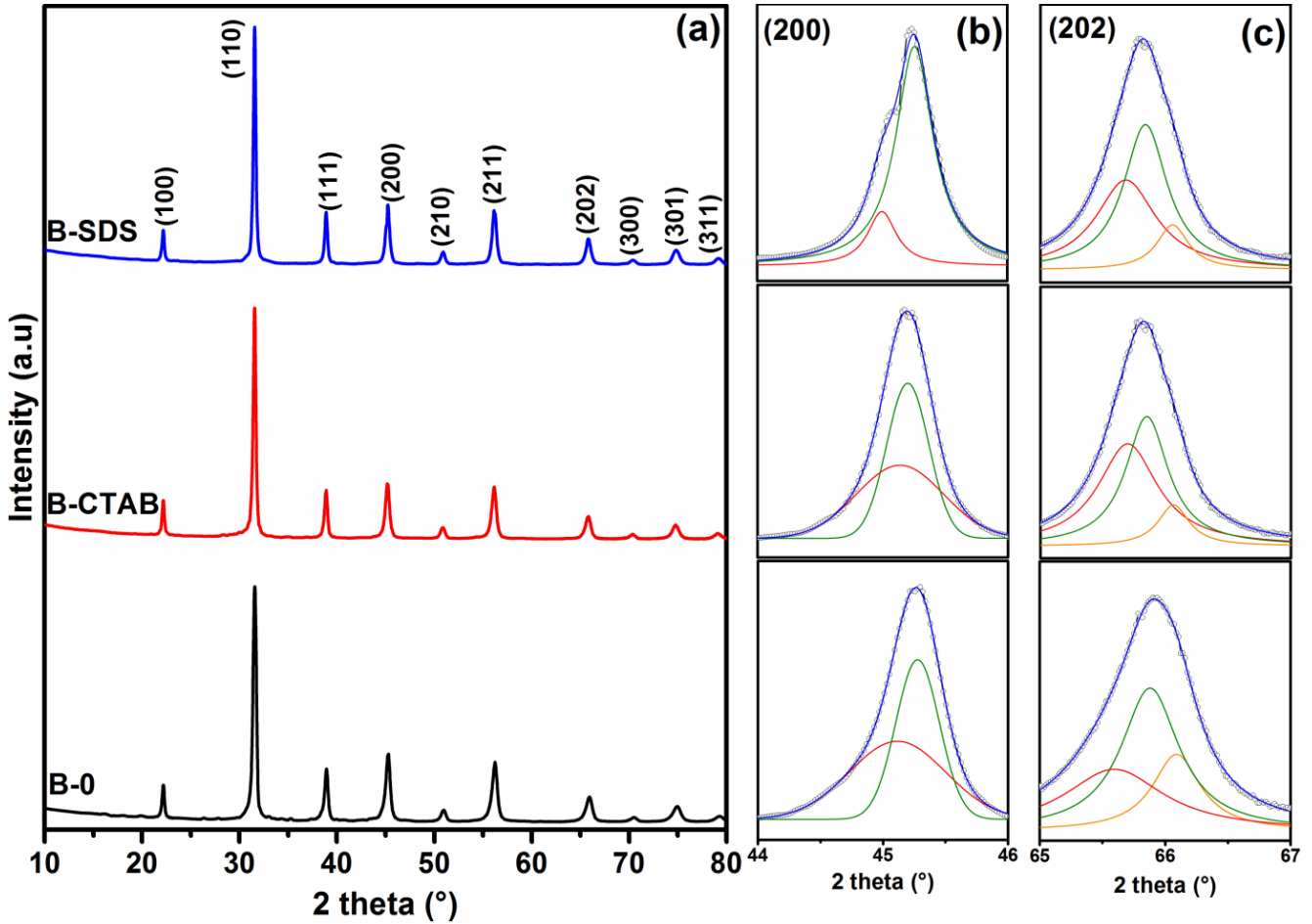

**Fig. 1.** (a) XRD patterns of BCZT powders calcined at 1000 °C/4h. (b) The enlarged views of the peaks 2θ ≈ (b) 45° and (c) 66°.

Phase analysis of the all BCZT powders was refined by considering Amm2 (orthorhombic) and P4mm (tetragonal) space groups. Table 1 summarizes the lattice parameters and structural information such as crystal space group and cell volume obtained after fitting using Amm2 and P4mm. Lattice parameters and space group are comparable to those reported in the literature [30–33]. The degree of tetragonality (c/a) of B-0, B-CTAB and B-SDS are 1.0013, 1.0036 and 1.0051, respectively. These values are in good agreement with those of Liu and Ren [12] and Chandrakala et al. [32]. The pronounced peak split in B-SDS powder was reflected in the high degree of tetragonality (c/a = 1.0051).

**Table 1.** Structural parameters obtained from Rietveld refinement using P4mm and Amm2 group spaces.

| Sample | B-0 | B-CTAB | B-SDS | B-0 | B-CTAB | B-SDS |
|---|---|---|---|---|---|---|
| **Space group** | P4mm | | | Amm2 | | |
| **Lattice parameters (Å)** | a = 4.0058<br>b = 4.0058<br>c = 4.0112 | a = 4.0017<br>b = 4.0017<br>c = 4.0162 | a = 3.9981<br>b = 3.9981<br>c = 4.0186 | a = 4.0214<br>b = 4.0429<br>c = 4.0158 | a = 4.0243<br>b = 4.0384<br>c = 4.0204 | a = 4.0211<br>b = 4.0430<br>c = 4.0162 |
| **Angles (°)** | α = β = γ = 90 | α = β = γ = 90 | α = β = γ = 90 | α = β = γ = 90 | α = β = γ = 90 | α = β = γ = 90 |
| **Volume (Å³)** | 64.3654 | 64.3138 | 65.1778 | 65.2895 | 65.3393 | 65.2934 |
| **Tetragonality (c/a)** | 1.0013 | 1.0036 | 1.0051 | - | - | - |
| $R_{wp}$ (%) | 11.60 | 8.18 | 6.81 | 15.3 | 14.9 | 16.1 |
| $R_p$ (%) | 10.60 | 8.30 | 6.96 | 14.8 | 13.9 | 11.8 |
| $R_{exp}$ (%) | 10.11 | 10.89 | 5.46 | 11.28 | 11.22 | 11.15 |
| $\chi^2$ | 1.31 | 0.56 | 1.55 | 1.84 | 1.76 | 2.08 |

*3.2. Microstructure of sintered ceramics*

Fig. 2 displays the SEM micrographs and grain size distributions of B-0, B-CTAB and BCZT-SDS ceramics sintered at 1250 °C/10h. The grain size distributions were determined using ImageJ software. B-0 sample exhibits a Gaussian distribution of grain size characterized by a mean grain size of 3.6 μm (Table 2). The grains are shown to be poly-faceted and the presence of inter-granular pores were highlighted. B-CTAB ceramic has a morphology relatively different to that observed in B-0. It is characterized by the presence of two types of grains: non-uniform poly-facets and relatively uniform spheroids (Fig. 2b). The grains are characterized by a lognormal distribution around the smallest sizes (around 1 μm). The resulting mean grain size was measured to be 2.7 μm (Table 2). It is worthy to mention that B-0 and B-CTAB have almost the same grain size but a different grain size distribution.

The porosity in B-CTAB ceramic is much lower compared to B-0. Indeed, the small spheroidal grains (60 – 1000 nm) were distributed at the boundaries of coarse grains and occupy the pores created during the sintering of B-CTAB sample. B-SDS ceramics exhibit a Gaussian distribution of grain size characterized by a mean grain size of 6.6 µm (Table 2). The grain boundaries are well defined and few facetted (Fig. 2c).

Finally, a high value of relative density was observed in B-SDS ceramics (96.4 %) followed by B-CTAB (95.2 %) and B-0 (93.1 %) (Table 2). This corroborates SEM results, where the distribution of pores decreases in B-CTAB and B-SDS ceramics.

One of the more significant findings to emerge from Fig. 2 is that the SDS surfactant allows an excellent control of the morphology of BCZT ceramic' grains during sintering.

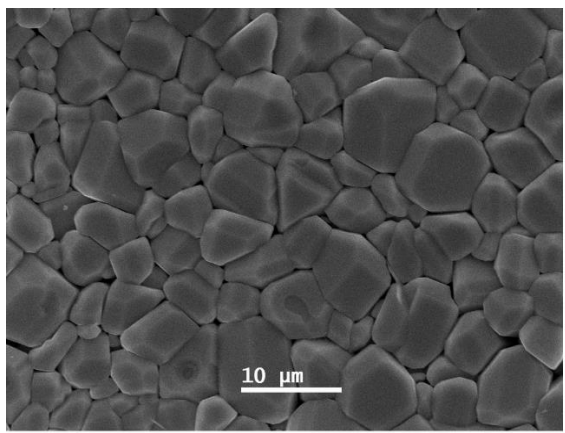
(a)
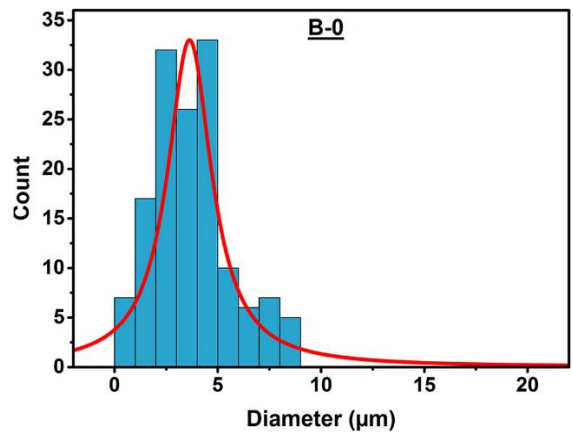
(d)
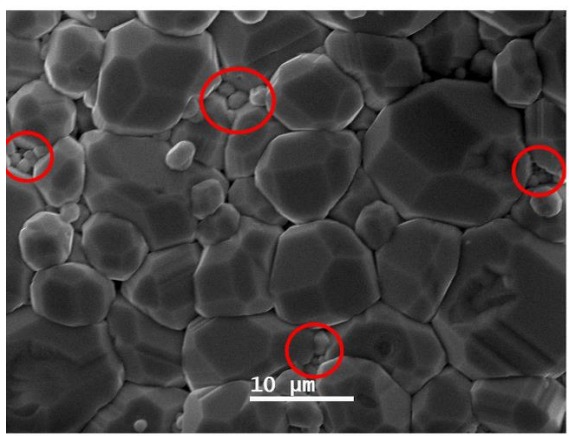
(b)
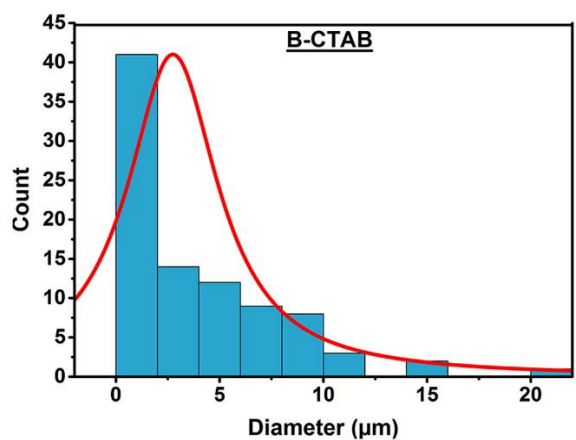
(e)
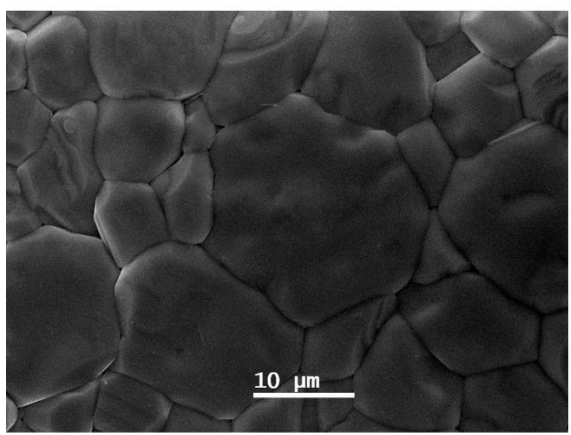
(c)
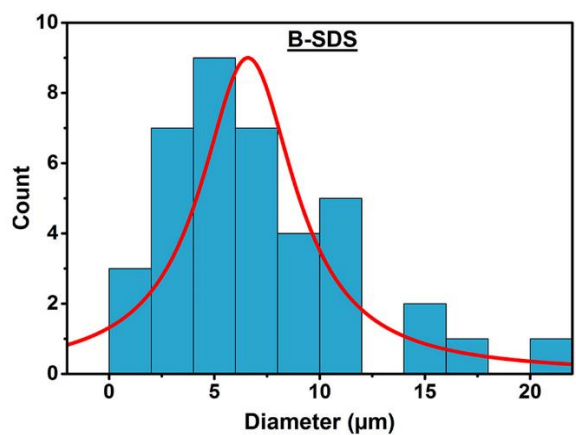
(f)

**Fig. 2.** SEM micrographs (a-c) and grain size distributions (d-f) of B-0, B-CTAB and B-SDS ceramics sintered at 1250 °C/10h.

**Table 2.** Relative densities and grain size of BCZT ceramics sintered at 1250 °C/10h.

|  | B-0 | B-CTAB | B-SDS |
|---|---|---|---|
| **Grain size (µm)** | 3.6 | 2.7 | 6.6 |
| **Relative Density (%)** | 93.1 | 95.2 | 96.4 |

*3.3. Dielectric properties of sintered ceramics*

Fig. 3 shows the temperature dependence of the dielectric constant at various frequencies for sintered BCZT pellets at 1250 °C/10h. The overall dielectric properties of all ceramics are summarized in Table 3. All samples exhibit a broad dielectric anomaly associated with tetragonal-cubic (T–C) phase transition around 90 °C. This dielectric broadening is due to compositional fluctuations and micro-inhomogeneity [34]. Meanwhile, all peak temperatures are dependent on the frequency and shift to high temperature with frequency increasing. In other words, all BCZT ceramics represent a relaxor behavior [35,36].

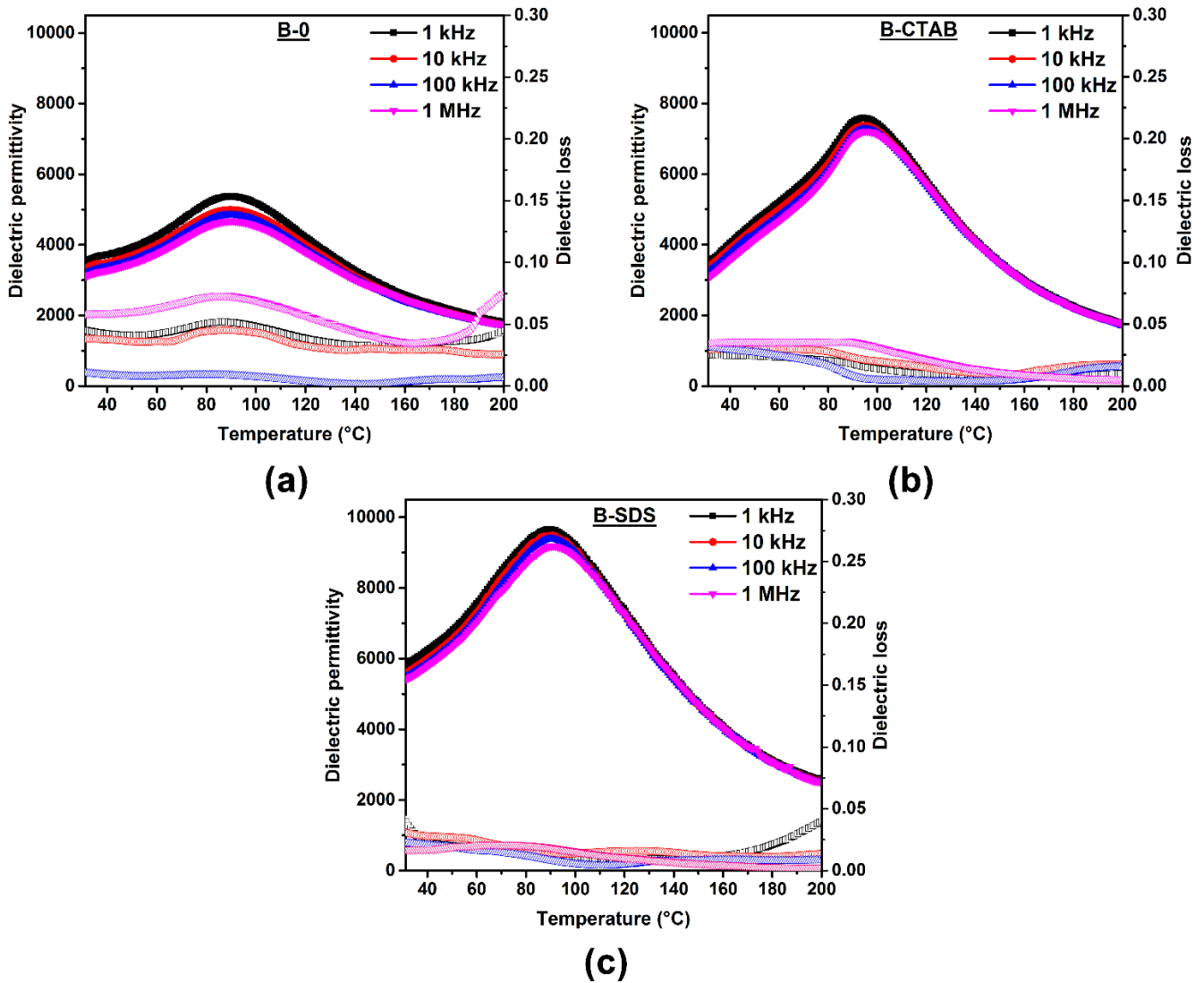

**Fig. 3.** Temperature dependence of dielectric permittivity and dielectric loss of (a) B-0, (b) B-CTAB and (c) B-SDS ceramics sintered at 1250 °C/10h.

Fig. 4 reveals the results of the deconvolution of peak temperature dependence of dielectric permittivity at 1 kHz of BCZT ceramics. It indicates the existence of two anomalies corresponding to orthorhombic-tetragonal (O-T) transition at room temperature (green curve), and a broad peak for tetragonal-cubic (T-C) phase transition at ~ 90 °C (red curve). These results bear out those of XRD suggesting the coexistence of orthorhombic and tetragonal phases at room temperature [18,30,37,38].

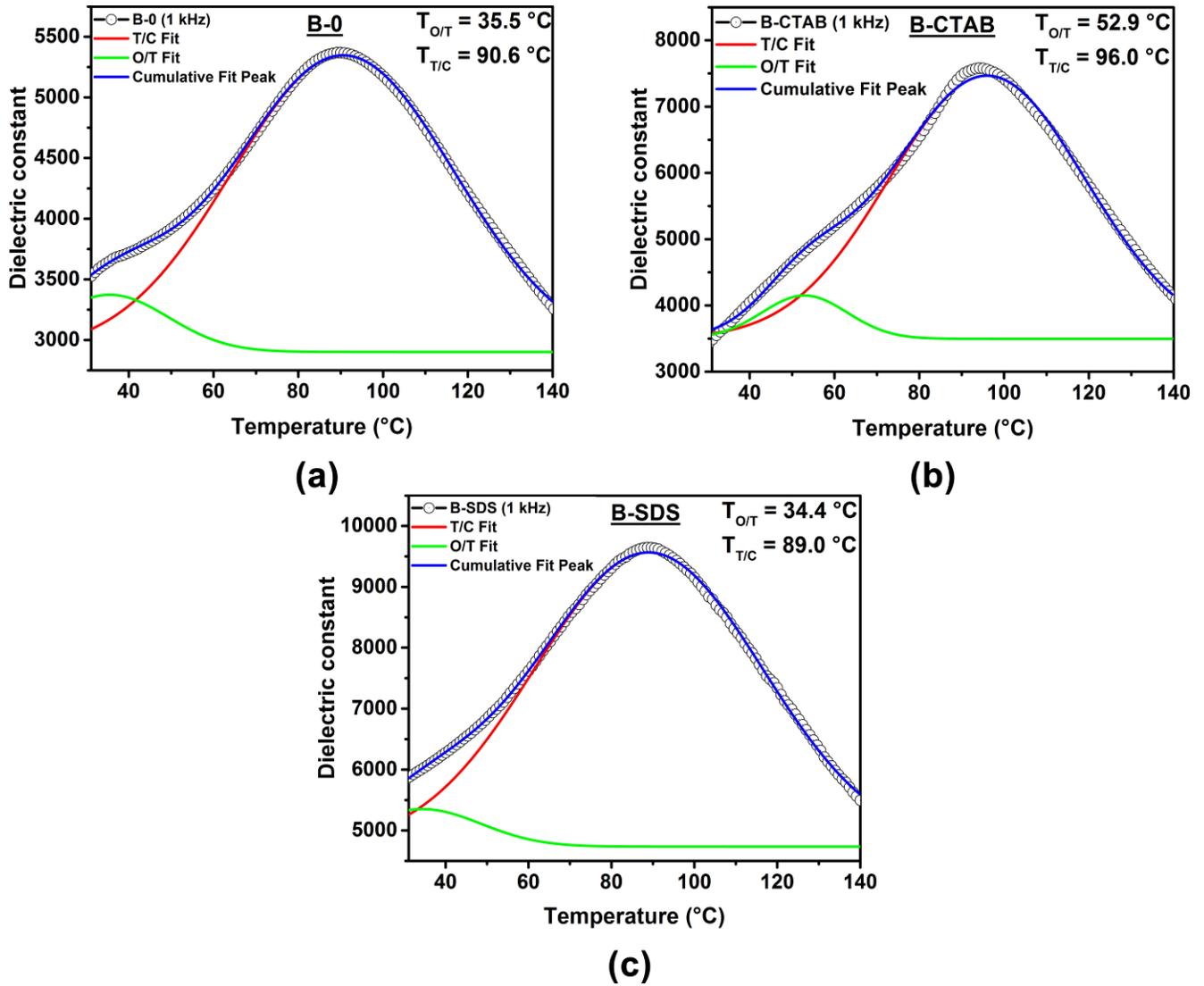

**Fig. 4.** Convolution of peak temperature dependence of dielectric permittivity at 1 kHz of (a) B-0, (b) B-CTAB and (c) B-SDS ceramics sintered at 1250 °C/10h.

To have an insight into this phase transition, the inverse dielectric constant was plotted as a function of temperature at 1 kHz and fitted using Curie–Weiss law:

$$\frac{1}{\varepsilon_r} = \frac{T - T_0}{C} (T > T_0) \qquad (1)$$

Where $\varepsilon_r$ is the real part of dielectric constant, $T_0$ is the Curie Weiss temperature and C is the Curie-Weiss constant.

The inverse of the dielectric constant as a function of temperature at 1 kHz for all BCZT ceramics is plotted in Fig. 5, and the fitting results obtained using Eq. (1) are listed in Table 3. It was found that $T_0$ changes from 67 °C to 100 and 94 °C as a result of the addition of CTAB and SDS, respectively. Besides, the real part of the dielectric constant ($\varepsilon_r$) of all samples deviates from the Curie–Weiss law above the Curie temperature. The deviations $\Delta T_m$ is defined by the following equation [14],

$$\Delta T_m(K) = T_{cw} - T_m \quad (2)$$

Here, $T_{cw}$ refers to the temperature from which $\varepsilon_r$ starts to follow the Curie–Weiss law, and $T_m$ denotes the temperature at which $\varepsilon_r$ value reaches the maximum. The calculated $\Delta T_m$ of B-0, B-CTAB and B-SDS ceramics are 37, 51 and 61 °C, respectively.

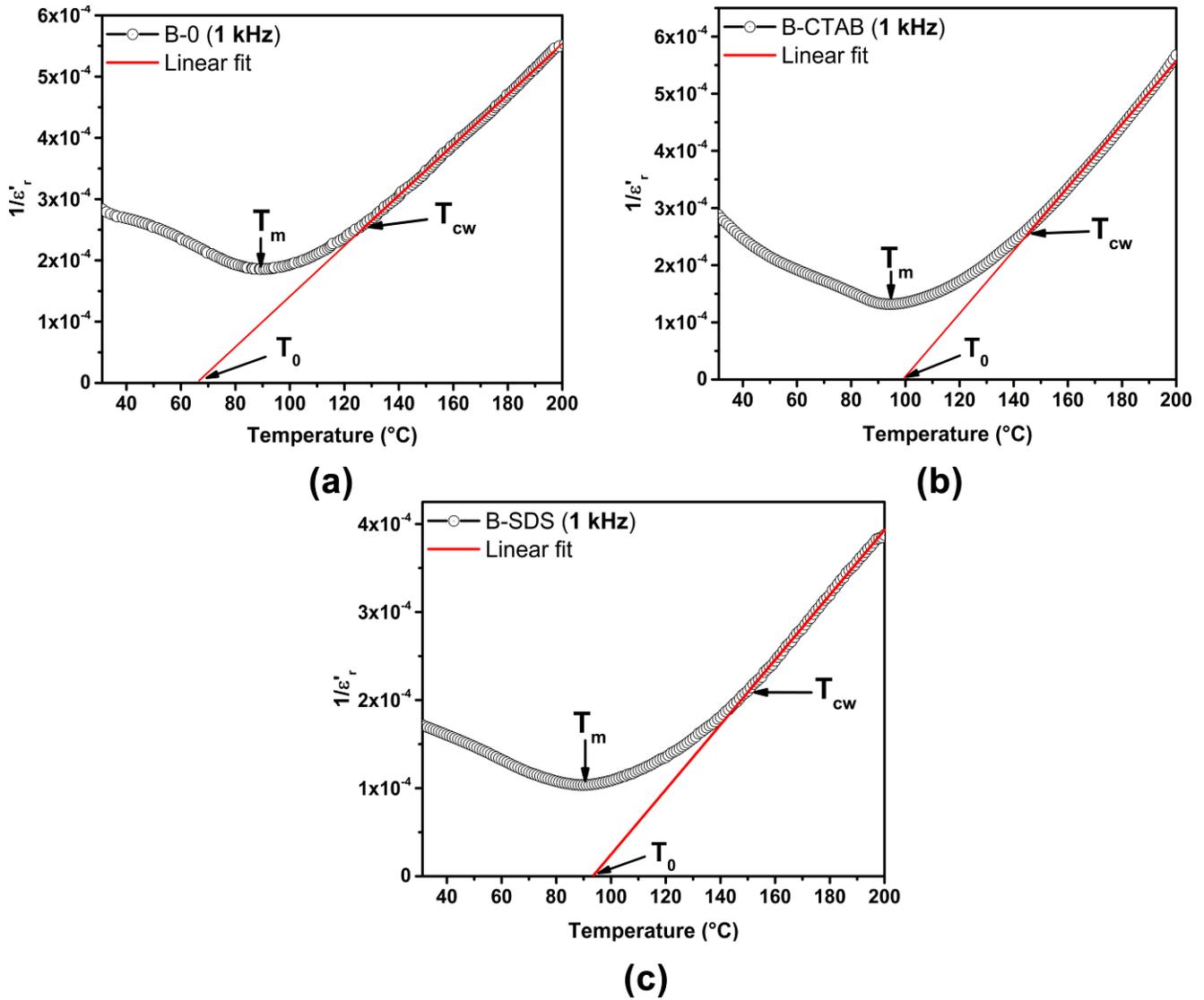

**Fig. 5.** Plots of Curie-Weiss relation for (a) B-0, (b) B-CTAB and (c) B-SDS ceramics sintered at 1250 °C/10h.

All ceramics are found to be associated with the Curie constant value of the order of $10^5$ K, which are consistent with that of typical well known displacive-type ferroelectric such as $BaTiO_3$ ($1.7 \times 10^5$ K) [39]. This indicates that high-temperature paraelectric phase is driven by a displacive phase transition. To elucidate the diffuseness associated with the transition, generally, Uchino and Nomura [40] empirical relation is employed to describe the variation of dielectric constant as a function of temperature above $T_c$ for relaxors. The relation is as follows:

$$\frac{1}{\varepsilon_r} - \frac{1}{\varepsilon_m} = \frac{(T - T_0)^\gamma}{C} (1 < \gamma < 2) \qquad (3)$$

where ε' max is the maximum value for the real part of the dielectric permittivity, and γ and C are constants. For an ideal relaxor ferroelectric γ =2, while for a normal ferroelectric γ =1 and the system follows the Curie-Weiss law [41]. Fig. 6 depicts the linear relationship between ln (1/$\varepsilon_r$ - 1/$\varepsilon_m$) and ln (T – $T_m$) at 1 kHz for all sintered ceramics. By curves fitting using Eq. (3), the values obtained for γ, determining the degree of diffuse transitions, at 1 kHz for B-0, B-CTAB and B-SDS ceramics are 1.775, 1.792 and 1.700, respectively. The high γ values in B-0 and B-CTAB ceramics indicate that the diffuse phase transition is supported by the small grain size [41,42]. In this context, enhanced relaxor behavior is obtained while decreasing grain size in BCZT ceramics [43,44].

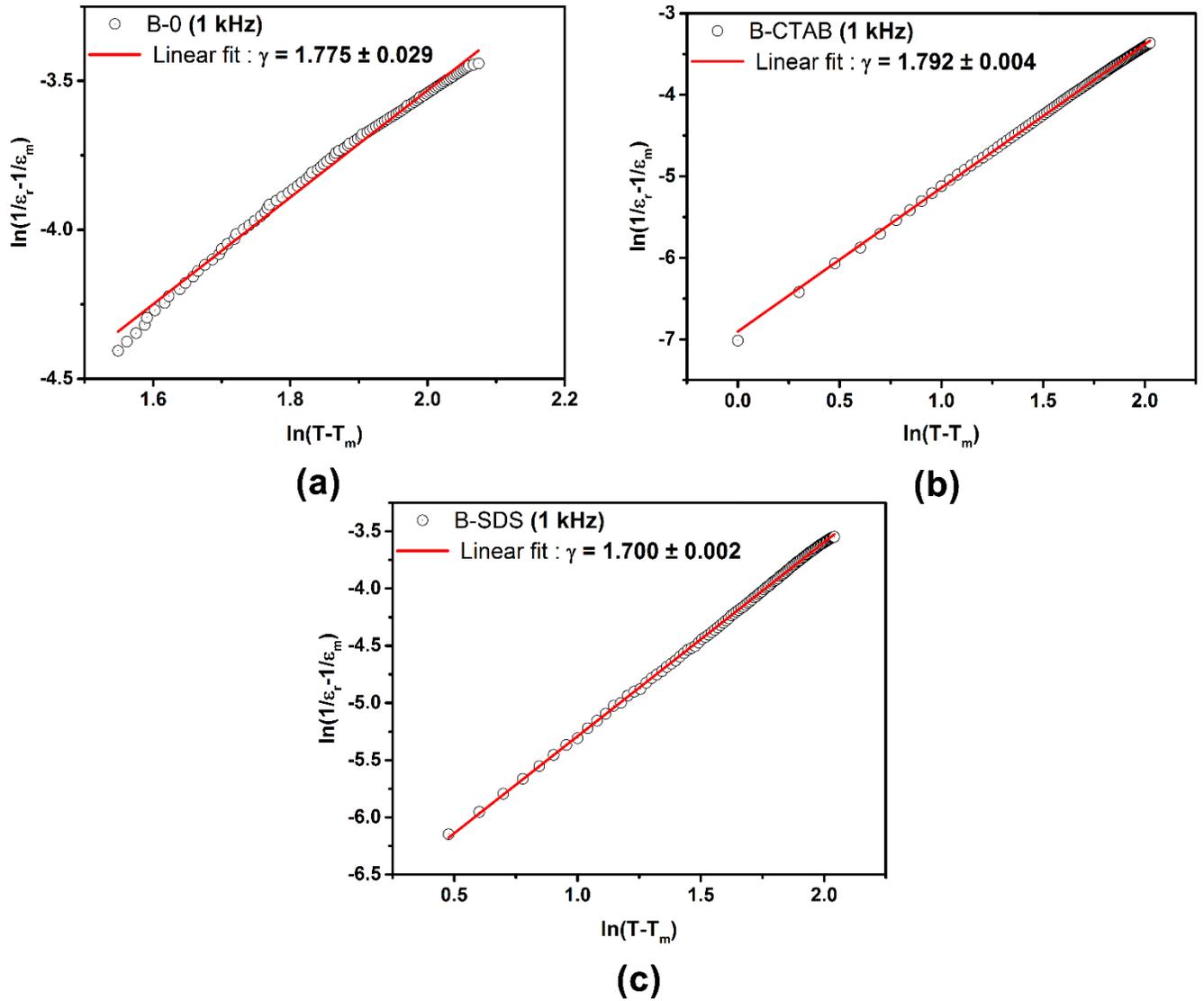

**Fig. 6.** Plots for modified Curie-Weiss law to determine slope (γ) for (a) B-0, (b) B-CTAB and (c) B-SDS ceramics sintered at 1250 °C/10h.

**Table 3.** Relaxor properties at 1 kHz of B-0, B-CTAB and B-SDS ceramics sintered at 1250 °C/10h.

|  | $\varepsilon_r$ | tan δ | Grain size (µm) | $T_0$ (°C) | $C \times 10^5$ (K) | $T_m$ (°C) | $T_{cw}$ (°C) | $\Delta T_m$ (°C) | γ |
|---|---|---|---|---|---|---|---|---|---|
| **B-0** | 5370 | 0.0517 | 3.6 | 67 | 2.429 | 89 | 126 | 37 | 1.775 |
| **B-CTAB** | 7584 | 0.0158 | 2.7 | 100 | 1.812 | 94 | 145 | 51 | 1.792 |
| **B-SDS** | 9646 | 0.0125 | 6.6 | 94 | 2.716 | 90 | 151 | 61 | 1.700 |

Traditionally, it has been argued that the increase in grain size of sintered ceramics enhances their dielectric properties [5,25]. As mentioned earlier, for B-SDS ceramics, the dielectric constant and dielectric loss are enhanced with 80% and 70%, respectively, after increasing the grain size from 3.6 to 6.6 µm. However, in B-CTAB ceramics, the increased dielectric constant was accompanied by a decreasing in grain size as described in Devonshire's phenomenological theory [45,46]. Moreover, it was believed that the increase in density contributed to the enhancement of $\varepsilon_r$ [19]. As revealed in densities measurements and SEM micrographs, the smaller particles could fill the gaps between the larger grains (Fig. 2b). This could result in increases in a degree of the packing and thus an enhancement of densification during the sintering process [47]. These suggest that the presence of surfactants exerts a favorable influence on the densification process. Consequently, enhanced dielectric properties even at small-grained BCZT ceramics.

Table 4 summarizes the dielectric properties, grain size average and conditions of synthesis of lead-free $Ba_{0.85}Ca_{0.15}Ti_{0.9}Zr_{0.1}O_3$ (BCZT) ceramic at 1 kHz. Our B-SDS ceramic with a small grain size exhibits a relatively improvement in its dielectric properties (the maximum of dielectric constant ($\varepsilon'_m$) and dielectric loss (tan δ)) compared to those cited in Table 4.

**Table 4.** Comparison of dielectric properties and microstructure reported here with other lead-free BCZT ceramics at different synthesis conditions at 1 kHz.

| $\varepsilon'_m$ | tan δ | Grain size (µm) | Synthesis conditions | | Ref. |
|---|---|---|---|---|---|
| | | | **Method** | **Sintering** | |
| 9646 | 0.0125 | 6.6 | Solvothermal | 1250 °C/10h | This work (B-SDS) |
| 7760 | 0.1000 | 12.09 | Hydrothermal | 1300 °C/3h | [41] |
| 5342 | 0.0255 | 6 - 10 | Sol-gel auto combustion | 1450 °C/2h | [45] |
| 4762 | 0.0220 | - | Solid-state | 1427 °C/2h | [31] |
| 17 375 | 0.0126 | 15.3 | Solid-state | 1300 °C/6h | [48] |

## 4. Conclusion

In conclusion, a novel strategy to reveal the effects of grain size distribution on dielectric properties of lead-free BCZT ceramics was reported. Three types of highly dense ceramics with different grain sizes and enhanced dielectric properties were successfully prepared at fixed low temperature sintering (1250 °C). Two mechanisms were proposed to explain the increase in the dielectric properties: (i) increase in grain size in coarse-grained ceramics (B-SDS) and (ii) enhancement of densification during the sintering process in small-grained ceramics through grains packing (B-CTAB). The proposed method elucidates that there is no special reason that the correlation of grain size effects and dielectric properties of lead-free or lead-based systems must be done by varying sintering temperature or dwell time. Indeed, the findings suggest that surfactant-assisted solvothermal processing is a powerful technic for grain size, shape and orientation controlling in ceramic nanopowders. Overall, it is worthy to note that increasing sintering temperature or dwell time could further improve the dielectric properties of the investigated ceramics. Based on the promising findings presented in this paper, it would be interesting to divulge the effect of microstructure engineering on electrical properties of these lead-free BCZT ceramics using complex impedance spectroscopy (CIS).

## Acknowledgements

The authors gratefully acknowledge the generous financial support of CNRST Priority Program PPR 15/2015 and H2020-MSCA-RISE-2017-ENGIMA action.